# ORGANIZATIONAL RESILIENCE BETWEEN COMPETING NETWORKS OF INFOMEDIARIES: A CASE STUDY IN CIVIL SOCIETY RESILIENCE IN HONG KONG

Sophie Zinser, United Nations University Institute in Macau, sophie.zinser01@gmail.com

Hannah Thinyane, United Nations University Institute in Macau, hannah@unu.edu

**Abstract:** This study explores how non-governmental organizations (NGOs) in Hong Kong can be considered as 'infomediaries' (UNDP, 2003) in their use of information and communication technologies (ICTs) to support resilience-building across a growing population of migrant domestic workers (MDWs). It also acknowledges MDWs effective existing self-organizing community networks, including religious groups and labour unions. This study maps how NGO infomediaries are currently supporting MDW communities. It posits that NGOs are uniquely capable of developing ICTs grounded in local legal, psychological, and cultural contexts to improve MDW community resilience. The study finds that the fragmented nature of technology use between NGO infomediaries and the competition between NGOs for funding hinders NGO infomediaries' ability to support building lasting resilience within the MDW community. Recommendations from this study seek to align NGO infomediary tool development more closely with the MDW community in Hong Kong's existing communicative ecologies. It considers NGOs as infomediaries capable of adapting and streamlining various linkages across grassroots MDW social organizations, local community leaders, and governments that impact the MDW community. This study is a tool for NGO infomediaries to understand the types of resilience networks that they are uniquely capable of building with MDWs in Hong Kong.

## 1 INTRODUCTION

COVID-19 has deeply impacted the global labour market, particularly when it comes to migrant domestic workers (MDWs). New estimates by the International Labour Organisation (ILO) reveal that 37% of female MDWs are at risk of unemployment ("Livelihoods," 2020). This is also particularly pronounced in Hong Kong, Special Administrative Region (SAR), home to a majority female demographic of over 390,000 MDWs from Philippines and Indonesia. Because Hong Kong's live-in rule stipulates that MDWs must live with their employers (GovHK, 2020), many MDWs who have recently been let go due to pandemic conditions currently lack shelter and other resources due to COVID-19. As a result, Hong Kong's dozens of non-governmental organizations (NGOs) who provide services to support this population's evolving needs are overwhelmed. Since January 2020, local headlines in Hong Kong reported testimonials from MDWs and NGO workers on difficult conditions facing MDWs including the acquisition of new debts, high recruitment agency and/or visa fees due to changed circumstances, and even struggles with homelessness (Siu & Bethoux, 2020).

This community's needs will only increase as social distancing and travel restrictions fluctuate. Therefore, an urgent communication gap has evolved between NGOs and the MDW communities in Hong Kong. Information communication technologies (ICTs) are increasingly being used to improve communication between the NGO sector and the populations that they serve. These tools do not themselves have agency, but can be developed and designed to empower worker balance of information asymmetries that exist within exploitative employer-employee relationships (Thinyane & Sassetti, 2020). But prominent gaps exist in understanding how NGOs as infomediaries can better use technology to access services and/or other forms of support in Hong Kong.

We understand labour exploitation as a spectrum, ranging from labour compliance at one end, through various labour and criminal law violations, to extreme exploitation or 'forced labour' at the other (Skrivankova, 2010). With this in mind, we can see that work situations that begin as consensual and mutually beneficial, can transform to oppressive and exploitative environments. These types of changes in work conditions can occur as a result of changes in personal (e.g. age), situational (e.g. employment type, migration status), or circumstantial (e.g. economic downturn) vulnerabilities. In addition, these types of vulnerabilities often reinforce and compound each other.

It is important to contextualize NGO infomediaries' engagements with MDWs in the context of the incredible self-organizing capacity of MDWs across multiple local grassroots organizations. In fact, since its coordinated actions for the anti-World Trade Organization (WTO) protests hosted in Hong Kong in 2005, the MDW community in Hong Kong is widely considered one of the most self-organized groups for grassroots action and activism (Bethoux, 2020). This is particularly true when compared to other MDW communities in Singapore, Taiwan, Malaysia, and the Gulf States. As a whole, the NGO infomediaries who are the subject of our study do not have broad-based engagement with these activist NGOs or labour unions. Instead, they exist instead inside their own insular discursive layer of NGOs most of whom receive funding from international organizations or local advocacy groups composed of mostly expats. The dissonance between communication between activist NGOs/union organizers and NGO levels in the MDW community in Hong Kong is an area demanding further study.

Studies have shown that information and communication technologies for development (ICT4D) can be designed to facilitate the social accountability necessary for the NGO sector to hold states accountable for services provided (Grandvoinnet, Aslam, & Raha, 2015). We are particularly interested in if and how NGO infomediaries for MDWs in Hong Kong can better develop ICTs grounded in local legal, psychological, and cultural contexts. In the conclusion section of this paper, we hope to suggest recommendations such that that any ICTs designed by NGO infomediaries are used to improve MDW decision-making and directly catalyse adaptive responses to protracted crises – across governments, NGOs, and local grassroots organizations.

## 2   THEORETICAL LINKAGES: RESILIENCE, COMMUNICATIVE ECOLOGIES, INFOMEDIARIES

The ability of global systems to "withstand, recover from, adapt to, and potentially transform amid external stressors," is referred to as resilience (Heeks & Ospina, 2019). Originally coined as a term to defining the capacity of ecological systems to withstand change, the term "resilience" has been found to be both 'abstract and malleable enough' to bring seemingly divergent fields and sectors – across security, finance, and development infrastructure – under the umbrella of a single metric (Walker & Cooper, 2011: 144). Resilience-building's importance is also predicated upon an increasingly turbulent global environment, in which populations must survive emergent such as climate change, pandemics, and financial crises.

Since the 2008 financial crisis and subsequent worsening of global inequality, discourse analysis on resilience reveals its growing prominence in the public policy sphere, particularly across the employment, health, and welfare sectors. (Allen & Bull 2018; Burman 2018) Subjects in need of resilience in the discourse are often lower-class women whose narratives centre around 'personal crises or accomplishments decoupled from economic and social circuits of accumulation and dispossession' (Gill & Ograd, 2018, p. 479). These vulnerable tales of struggle are contrasted with a group of imagined mostly middle-class Western women who are able to draw on networks of information, psychological, and financial support to actualize resilience in an increasingly chaos-defined world (Gill, 2018). Idealized subjects' success in the context of resilience frameworks rely on constructive relational contexts, in which communication and information-based resources are vital steps towards improving their circumstances. (Jefferis & Teron, 2018). We refer to the agents operating within these constructive relational contexts as "infomediaries," – short for information intermediaries – who "synthesise, translate, simplify and direct information on behalf of others." (UNDP, 2003). In the context of MDW communities in Hong Kong, we consider NGOs to be necessary infomediaries because they are bridging gaps between MDW communities, decision makers in government, and local legal and social services systems. Infomediaries here act as the missing link between those in need of building resilience and the idealized scenario of groups meeting resilience objectives.

Literature on the relationship of information communication technology for development (ICT4D) to resilience situates the concept in a broader moral and social context of the societies that it references (Heeks & Ospina 2019). This uniquely offers alternatives to other streams of resilience literature that largely focus on recovery and continuity of vulnerable communities because it suggests that tools can be designed enabling communities to adapt to different methods of resilience. New avenues for a critical perspective on resilience can therefore be found in ICT4D studies that identify the need for increased adaptation and solution-focused approaches (Chen, 2015). Without these emphases on adaptation, communities may remain resiliently poor, or alternatively inequitable societies will not sustain in the long-term (Marais, 2015:436, Wilkinson & Pickett, 2010)

Hong Kong's MDW community is a fitting case study for resilience as it has also been the recipient of rapid patterns of technology diffusion. Specifically, internet plans for mobile phones have become cheaper, as public Wi-Fi uptake – particularly across Asia – has increased alongside a nearly universal adoption of social media as a communication tool between social groups and across diverse socioeconomic classes, race, gender, and national backgrounds. ICTs have become the go-to sources for information and data that drives individual decision-making processes – regarding everything from reporting data to sharing tips on arrival and recruitment processes for workers (Heeks & Ospina 2019). The sources and credibility of the information that MDWs are exposed to on social media to inform their decision making is a significant gap that demands further empirical study. Any study of how use of ICTs impact MDW communities in Hong Kong should take into account the depths of existing in-person and digital relational networks. Multi-stakeholder roles in technology solutions can be best approached through an ecosystem perspective, because all interactions between relevant stakeholders need to be considered when mapping the most effective pathways for solutions (Thinyane & Goldkind, 2018).

To interrogate how infomediaries operate in Hong Kong, it is important to consider the existing systems that perpetuate the spread of information within Hong Kong between NGOs, faith-based organizations, individual MDWs, MDW community, labour associations , and local government officials. Just as resilience studies have their origins in ecology, Tacchi, Slater and Hearn's (2003) definition of communicative ecologies is another helpful model for holistically perceiving of discourse by, about, and for MDWs in Hong Kong. Communicative ecology is a conceptual model used to represent the relationships between layers of discourse, social interactions (whether people or groups of people), and technology within a particular community. It relies on an ecological metaphor to understand the "processes that involve a mix of media, organized in specific ways, through which people connect with their social networks" (2003, p. 17). More than just a mapping of which platforms NGOs, MDWs, and government officials use, this approach addresses the nuance of how patterns and processes of communication already occur. Subsequently, a communicative ecology interrogates the value and worth of such communication networks based on the 'levels of strata' or activity in which they already inhabit (Thinyane & Siebörger 2017). These technological levels are broad, encompassing everything from face-to-face communication (no-tech) and traditional media (television, radio – low tech) approaches, to more high-tech approaches including social media and customer relationship management (CRM) data management systems. Levels of how and why groups communicate with each other can be diverse, change overtime, and impact each other.  Interactions across these technological layers can occur across different types of technology reception and input: one to many, one on one, many to many, and more horizontal, peer to peer communication (Foth & Hearn, 2007, p. 9). The social layer consists of  the ways that the people are organized (informal social networks, associations, communities, legal entities). The discursive layer consists of the content that is discussed between different groups of people, using different technologies. Further, just as external social factors or patterns can influence technology usage, technology can impact existing discursive and/or social patterns (Hearn et al., 2014, p. 8). This conceptual model allows the researcher to frame communication of MDW infomediaries in the context of existing complex and nuanced communication structures.

Over the course of this study, we interrogated NGO infomediary-designed or developed ICT solutions to build (or impede) resilience for the growing, diverse community of MDWs in Hong Kong. Understanding resilience in the context of communicative ecologies in which MDWs are already participating encourages bottom-up decision making and increasing transparency between MDWs, and the NGOs/governments that serve them.

## 3　METHODOLOGY

As this research is intended as an exploratory study, we did not aim to seek a representative sample. We did not interview MDWs themselves due to restrictions on interviews and difficulties in accessing research interviews with MDWs posed by COVID-19 conditions. We also focused exclusively on the NGO community because our interest was in focusing the research on ICT tool development opportunities to build resilience for NGOs working with Hong Kong's MDW community. As was previously mentioned, too often the burden of resilience building falls directly on the community in need of resilience themselves. This study instead gives us a rich picture not of what MDWs think about their own communication models but how NGOs working closely with MDWs perceive that discourse. That being said, a glaring gap in this study remains the lack of input from MDWs themselves; an area necessary for further study once COVID-19 conditions and therefore interview access improves.

From November-December 2020, the [Anonymised] Research Team conducted interviews with 23 NGO workers from 14 NGOs. During the period of initial consultation, the IOM's Hong Kong Sub-Office acted as a mediating agency between researchers and local stakeholders. IOM wielded its integral local network, referring researchers to a wide range of direct and indirect stakeholders working on or around issues relating to MDWs. Once this initial round of interviews was completed in late November 2020, contacts were obtained through a snowball sampling method from participants. To anonymise responses, each NGO has been assigned a letter and a unique number referring to each participant from each organization.  As an example, two respondents from Organization A participated in interviews.  Quotations from the first respondent will be referred to as A1 and quotations from the second respondent will be referred to as A2 in the remainder of this document. Table 1 summarizes key details for each of the organizations involved in this study.  As this table indicates, participating organizations worked either: exclusively with MDWs; or with a combination of MDWs, ethnic minorities in Hong Kong, refugees and/or other marginalized groups as their beneficiaries.

| Organization Code | Number of Participants | Description |
|---|---|---|
| A | 2 | NGO offering legal and social support to ethnic minorities in Hong Kong, priority focus on MDWs |
| B | 4 | Organization supporting pregnant MDWs in Hong Kong |
| C | 2 | Think Tank / advocacy organization offering a range of programs to end human trafficking and forced labour in Hong Kong |
| D | 1 | Organization supporting with MDWs' employers in Hong Kong |
| E | 1 | Organization protecting the rights of MDWs and ethnic minority groups in Hong Kong |
| F | 2 | Organization working to support the legal rights of migrants in Asia, develops preventative solutions to MDW issues to enlist MDWs in validating their experiences with corrupt employment agencies |
| G | 1 | Legal rights and advocacy organization working with marginalized populations in Hong Kong |
| H | 2 | Organization working to empower migrant domestic workers in Hong Kong via online training and peer support |
| I | 3 | Advocacy organization that works to provide legal support to refugees and migrants in Hong Kong |
| J | 1 | Lead volunteer at local community shelter for MDWs affiliated with NGO A, academic researching MDWs in Hong Kong |
| K | 1 | Domestic helper agency supporting more ethical hiring of domestic workers in Hong Kong |
| L | 1 | Works with other NGOs and social enterprises that work with MDWs to identify their needs and connect them with pro-bono lawyer networks |
| M | 1 | Offers financial literacy training, support, and resources to MDWs in Hong Kong |
| N | 1 | Organization that provides funding and legal support to other organizations |

**Table 1: Participating organizations**

Interviews followed a semi-structured format, with eight interview questions, that shaped the direction of a semi-structured conversation. These questions covered participant perspectives on urgent needs facing migrant domestic workers in Hong Kong, current patterns of access to and use of technology, and the types of case-based support that MDW clients were accessing already. Interviews were transcribed and underwent three rounds of inductive coding using Nvivo to track and uncover results. Themes that were uncovered are presented in the next section to map the current resilience capacity for MDWs in HK from the NGO perspective.

# 4  RESULTS

This study provides an opportunity to understand how NGO infomediaries could re-design or improve existing ICT usage and communication patterns internally, across other NGOs, and to engage with the populations that they serve; focusing on an adaptive approach to building resilience. It assumes that NGOs will remain distanced from other types of grassroots worker associations that work with MDWs and thus does not claim that technology advances can alter divides that exist in current communicative ecologies surrounding MDWs in Hong Kong.

## 4.1  Context: MDWs Face a Deeply Oppressive Culture in Hong Kong

All 14 interviews contained vivid descriptions of physical, financial, or emotional mistreatment of MDWs at the hands of employment agencies, money lenders, the Hong Kong government, other NGOs, employers, or a combination therein. One participant who supports MDWs in situations of labour exploitation and forced labour described this oppressive treatment as:

> *…There is this sort of feeling that they [MDWs] must have done something wrong. When in reality, it's the system or employer that has done something wrong.* [G1]

From this comment it is clear that anti-immigrant racism, perceptions of MDWs as 'victims' rather than as professional employees, and decades of mistreatment have created a homogenously oppressive culture for MDWs in Hong Kong.

Employers' powers of access and control extended to MDWs' use of technology in Hong Kong. Withholding of ICT tools became a prominent outlet for employers to exert power and control.

> *Some employers are not happy because they think it distracts domestic workers from their work. That's why some ask the domestic workers to give their phone to them* [E1]

The taboo and restrictive culture around phone usage in what is both a workplace and the home is leading to confusion of MDWs, even when they need to use their phone for work.

> *Sometimes the workers find it difficult to ask questions of their employer like "When can I use my phone? There is something that I want to read up on online pertaining to my job, when can I do that?" Workers don't feel good about asking their employer those questions. How they access tech in their workplace depends on employer's personality and preferences* [D1]

The above examples remind us that developing ICT tools for NGOs to improve MDW resilience in Hong Kong must acknowledge the pervasively oppressive culture those workers face in their workplace daily.

## 4.2  Widely Held Perception that More Data will Lead to Better Policy/Programs for MDWs

Asked to describe the type of problems faced at work, NGOs pointed to the need for more nuanced data types about the population of MDWs currently in Hong Kong from the government side. Statistics needed included the current number of MDWs in Hong Kong grouped by origin country (A1,C3,G1,J1), the number of currently and previously detained MDWs (E1, I1, L1), the number of MDWs who have lodged complaints of sexual harassment against their employers (B3, I1 and L1), and the number of MDWs who currently have cases that have yet to be resolved in court (I1 and L1). It was unclear whether NGOs were not able to access this information because they had not filed a data information request to necessary government or immigration officials, or because the government was slow or inactive when it came to sharing data. NGO K claimed that they had submitted many government requests but had not received a proper response. NGO F bemoaned the inability of peer NGOs to lodge appropriate government requests, insisting that the information was readily available. NGO A claimed that they and their team had not thought to file an information request from the government and remarked that they would look into how to do that. The various opinions on how easily accessible this data was reflected different approaches across these NGOs and offers another example of a fragmented information sharing across the sector.

A1, F2, J1, M1 all mentioned how a gap in available data could result from the fact that MDWs are often hesitant to report situations of money laundering or lodge complaints due to long waits for trial times, or otherwise feared speaking up in defence of their rights. One interview with NGO J commented:

> *The hesitance comes from a sense of helplessness because of all of the structural constraints. These women are very traumatized, and most of them just have to continue, don't have a choice, only way they can support their children and families is to go again….They can work with NGOs, but will take a long time before anything changes.* [J1]

These gaps were further exacerbated by the racism that MDWs face often when coming forward. A representative from legal NGO G commented that:

> *The overt racism that we have experienced in tribunals in lower courts here in government is absolutely shocking. I remember going to tribunals and our clients were shouted at by the tribunal officers.* [G1]

MDW hesitancy in reporting may stem from justified fears of overt aggression and racism from the Hong Kong government.

Others mentioned that the burden of designing data-sharing systems should come from Hong Kong's NGOs and civil society themselves. One representative from NGO E had specific recommendations about how NGOs could design a system to identify bad employers. She asked:

> *How do we know that this employer is good? Sometimes we don't know. So then the agency will not pass the information if there is a rape case that is being lodged, we [the NGOs]will not know this information…[because] there is a data protection. How can this information be available or if it's possible so that the domestic workers can decide that they don't want to work here.* [E1]

K1, D1 and E1 – all representatives of NGOs that work closely with employers – mentioned that without designing such a system, oppressive employers can continue to operate in the MDW community, which can have disastrously negative impacts.

### 4.3   Rich Communicative Networks Already Exist Among MDWs

All NGO worker respondents noted that the primary tool that they used to communicate with MDWs was Facebook, either through messenger or group chats. Beyond traditional communication metrics, NGOs had different and often novel ways to make use of this platform. NGO H, which works with MDWs and online education, spoke about how Facebook maintains communication and a sense of community with alumni of their courses. A representative NGO C referenced the difficulties in communicating with MDWs and their families back home as health or psychological issues arose in Hong Kong. A third NGO B that works closely with MDWs who are pregnant spoke about how even facilitating social spaces for activities such as dancing, singing karaoke, or gameplay online. NGO F that works on preventative solutions to MDW issues Facebook to enlist MDWs in validating their experiences with corrupt employment agencies said:

> *So we will post content, or maybe they will post us a tip…[for] a suspicious agency on the street. We will post the photo being like "does anyone know this agency?" and then people* *will comment, usually within like 20 minutes. [They will say] Oh that's my agency, I used that five years ago, three years ago, two years ago. This sort of engagement is a lot less intimidating than downloading an app.* [F2]

In this example, NGOs were able to effectively act as infomediaries through engaging with MDWs using technology that they already use.

### 4.4 Solution Building Exercise Reflects Fragmented and Competitive Nature Amongst NGOs working with MDWs

As a whole, NGO staff gave mixed reviews on the potential of mobile apps to support MDW's access to information on either government or NGO services available in response to issues that may arise. While over half of surveyed organizations remained hopeful on the potential success of app-based solutions to aggregate data on MDWs in HK, many also cited fragmentation of service provision across NGOs and prior failed attempts to centralize information as issues around long-term uptake of such tools.

One worker with NGO E who personally had decades of work experience across several organizations that supported MDWs remarked that this has been a long-standing conversation across the sector:

> *In Hong Kong, a lot of us have talked about having one place where domestic workers can find everything. There are bits of things everywhere, but there should be one tech portal in any form, one tech portal in terms of where they could access everything that they need.* [E2]

Another representative from NGO A, a faith-based organization that works closely with MDWs suggested:

> *There is a lot of bits of information out there. We have FAQs… for legal issues and their rights. There are a lot of information all over the place, bring all of that together in one easily accessible place. There is a lot of knowledge and experience and expertise, but it is all very disparate.* [A1]

Given the generalist nature with which this solution was proposed, it was assumed that no single NGO had either a comprehensive understanding of the necessary information to build this database nor the resources to build something that centralized all data into one place. Further, representatives from all three of the aforementioned NGOs [A1, B1, E2] expressed that it would be difficult to find collaborative funding for such a project, and seven NGOs total [A1, B1, E2, F1, I1, L1, M1] – half of total respondents – made remarks regarding the "competitive" and/or "limited" nature of funding for the NGO sector in Hong Kong. Although not always drawn as a direct correlation, perhaps in some ways this fragmentation has developed as a bulwark towards progress across the sector.

### 4.5 The Importance of "Proactive" or "Preventative" Tech Tools and Solutions

Remarkably, responses that mentioned the importance of "proactive" or "preventative" solutions were discussed across all 14 interviews. In one section of the interviews, researchers brainstormed with respondents about potential pathways for ICTs to support MDWs in Hong Kong. Representatives from five NGOs [C2, D1, F1, F2, G1, L1] identified a need for more "proactive" or "preventative" solutions to issues facing MDWs in Hong Kong. NGO C remarked that:

> *One thing that collectively in civil society and government can do is be more responsive to the environment under which foreign domestic workers operate. For example, for the COVID-19 situation, how can improvements now not be playing catch up but be pre-*

> *emptive or have a faster response rate to meeting some of the needs of the workers that are coming to us.* [C2]

NGOs also had varying ideas regarding what "proactive" solutions really meant; that is how they could build ICT tools that would proactively build MDW resilience. As it is often expensive for MDWs to stay in Hong Kong without employment during an ongoing court case, NGO E, a legal organization that works with MDWs, suggested that:

> *Many financially need to drop their court cases in HK… then maybe allowing them to continue their cases through video interviews through not being in HK would help a lot of our clients. Especially when in COVID-19 they can't travel to come back for a police interview.* [E1]

NGO D that worked with employers of MDWs suggested providing MDWs and their employers with standardized yet culturally- and linguistically specific training to better manage MDWs or better understand the job responsibilities of being a MDW in Hong Kong. NGO G that had worked with MDWs on identifying harmful employment agencies suggested creating an ICT tool in the form of a privacy-preserving data repository for aggregating legal cases to be put forward to stop financial fraud in its tracks. Others suggested that NGOs may have the potential to design tools that would give a more broad-based view of the gaps in technology literacy for MDWs before they arrive in Hong Kong.

## 5  DISCUSSION

This work builds on existing research on the critical and primary role of filling gaps in local social services and mobilizing local communities (Bopp, Harmon, & Voida., 2017; Zhang & Gutierrez, 2007; Thinyane & Goldkind, 2018). In doing so, it identifies significant potential for the nuanced insights from the NGOs described in this study to be leveraged and integrated into resilience-building initiatives at a policy level, whether through Hong Kong's local government or a broader set of multilateral actors. Participants noted that MDWs are increasingly reliant on social media platforms (Facebook and WhatsApp) to communicate with one another, create community within Hong Kong, and exchange information on everything from their daily life to financial issues. To increase their efficacy, it became clear that there is an urgent need for NGOs to work with MDWs in Hong Kong to refer vulnerable cases among peer networks and identify issues as they arise. However, while Section 4.4 revealed a strong emphasis on data collection from MDWs, the risks and limitations of NGOs to collect data points given emerging data privacy concerns and MDW comfort with giving information over to NGOs based on socio-cultural perceptions of data as "personal information" remained unaddressed in our interviews, and perhaps an area of further study.

Many respondents remarked that significant gaps in the way that their funding streams were structured limited the widespread applicability of their programs across the entire MDW community and thus the efficacy of their impact. While fragmentation within the sector is endemic to civil society organizations globally, we believe it is exacerbated in the Hong Kong context. Many participants expressed exhaustion at combatting the systemic social issues that had existed since the original arrival of MDWs to Hong Kong from Indonesia and the Philippines throughout the 1970s. The purpose of each individual NGO's services did come across as diverse and therefore potentially symbiotic for collaboration with those of other organizations. For example, one targeted only pregnant MDWs, one focused on providing training for MDWs in financial literacy. But in practice, particularly regarding how these organizations described working with or for one another, they seemed to be competing for some of the same funding or limited funding streams, all to meet the needs of a small sub-section of Hong Kong's MDW population. Whether access to increased funding could lead NGOs closer towards effectively support a larger part of that population, or if large segments of Hong Kong's MDW population do not require NGOs' specialized services, was unclear given the current level of analysis. Given the findings in Section 4.4, one area for future research based on this study's findings would be the interactions between NGOs and MDWs and funders with long-term resilience in mind. The nature of the communication between the donors and the "beneficiaries" themselves is an area that likely needs improvement, and could be discussed in future studies.

# 6   CONCLUSION

Through the recommendations made by NGOs, this study shows the ways in which Hong Kong's NGO community that works with its MDW population is uniquely capable of designing ICT use for integration with existing platforms grounded in local contexts that might improve MDW the community's resilience.

As resilience-building is only now becoming more of a focus for the UN's international development agenda (IDA 2019), leveraging our understanding of how low-income mostly female communities like MDWs use technology will become increasingly integral to strengthening the levels or strata of resilience for both MDWs and their infomediaries. But using ICTs to improve MDW communities' resilience in the face of external shocks does nothing to address the root causes of those shocks in the first place. Given Section 4.3's discussion of MDWs existing self-organizing support networks, it also treads a fine line of moving the needle on responsibility for non-resilience closer to the communities of MDWs themselves, rather than the aforementioned social, cultural, and legal systems that perpetuate their vulnerability in Hong Kong. Therefore, NGO capacity to offer adaptive solutions to resilience in this space remains promising – as is evidenced in Section 4.2 and 4.5 – yet long-term capacity for NGOs to enact broad based change alone is nearly impossible given the complexity of the problem. However, with increased access to MDWs, NGO actors have the potential to make meaningful contributions to improving resilience for MDWs in Hong Kong. Through the lens of communicative ecologies, we in pointed out in Section 2 how MDWs are deeply engrained in the systems and networks of their community, which are often monitored and engaged with closely through NGO infomediaries.

Perhaps some of the dissonance in engaging across sectors to build resilience – particularly when competitive self-organization is involved – stems from confusion around how international development concepts such as resilience can actually support communities to better withstand existing shocks. SDG 8 for example – to "promote sustained, inclusive and sustainable economic growth, full and productive employment and decent work for all" – does not recognize the role of civil society actors towards promoting decent work. Nor does it provide clear pathways or guidelines for meaningful engagement with the diverse cross section of stakeholders with whom the SDGs' indicators necessitate engagement (Thinyane & Goldkind 2018). NGOs that work closely with MDWs were found to be integral to the increased direct participation of those workers towards brokering increased collaboration for social indicators. However, without alterations to the socio-legal system that build MDW communities' resilience, NGOs – while armed with the right tools – will not be able to realise the full potential of their contributions to the wider MDW community.

When considered as artifacts that when empathetically designed can rapidly adapt to solve nuanced problems, we assume that technology can be uniquely designed to improve communication across changing conditions. (Kumar, Karusala, Ismail, et al., 2019; Dombrowski, 2016: 16). But significant competition amongst those networks can prove a bulwark towards more effective attempts at communicating and exchanging ideas. Therefore, the focus of tools developed going forward should remain rooted in workers' experience without implying a possible failure of MDWs themselves to build resilience. Tools should instead highlight the positive potential of NGO infomediaries to work across levels of social organizations to affect change through – as was referenced numerous times across the study – developing proactive, preventative solutions. This will sustain resilience through providing information not just amongst members of diverse MDW communities but transmit that information upwards to local government and multilateral organizations.